\documentclass[aps,prl,reprint,showpacs,superscriptaddress]{revtex4-2}
\usepackage{graphicx}
\usepackage{dcolumn}
\usepackage{bm}
\usepackage{color}
\usepackage{lipsum}
\usepackage{mathrsfs}
\usepackage{lineno}
\usepackage[colorlinks,urlcolor=blue,linkcolor=blue,citecolor=blue]{hyperref}
\usepackage{siunitx}
\begin{document}


\title{Transport of intense ion beams in plasmas: collimation and energy-loss reduction}
\author{Yongtao Zhao}
\email{Email: zhaoyongtao@xjtu.edu.cn}
\affiliation{MOE Key Laboratory for Nonequilibrium Synthesis and Modulation of Condensed Matter, School of Physics, Xi'an Jiaotong University, Xi'an 710049, China}
\author{Benzheng Chen}
\affiliation{MOE Key Laboratory for Nonequilibrium Synthesis and Modulation of Condensed Matter, School of Physics, Xi'an Jiaotong University, Xi'an 710049, China}
\affiliation{Institute for Fusion Theory and Simulation, Department of Physics, Zhejiang University, Hangzhou 310058, China}
\author{Dong Wu}
\email{Email: dwu.phys@zju.edu.cn}
\affiliation{Institute for Fusion Theory and Simulation, Department of Physics, Zhejiang University, Hangzhou 310058, China}
\author{Rui Cheng}
\email{Email: chengrui@impcas.ac.cn}
\affiliation{Institute of Modern Physics, Chinese Academy of Sciences, Lanzhou 730000, China}
\author{Xianming Zhou} 
\affiliation{MOE Key Laboratory for Nonequilibrium Synthesis and Modulation of Condensed Matter, School of Physics, Xi'an Jiaotong University, Xi'an 710049, China}
\affiliation{College of Physics and Electronic Engineering, Xianyang Normal University, Xianyang 712000, China}
\author{Yu Lei} 
\affiliation{Institute of Modern Physics, Chinese Academy of Sciences, Lanzhou 730000, China}
\author{Yuyu Wang} 
\affiliation{Institute of Modern Physics, Chinese Academy of Sciences, Lanzhou 730000, China}
\author{Xin Qi} 
\affiliation{Institute of Modern Physics, Chinese Academy of Sciences, Lanzhou 730000, China}
\author{Guoqing Xiao}
\affiliation{Institute of Modern Physics, Chinese Academy of Sciences, Lanzhou 730000, China}
\author{Jieru Ren} 
\affiliation{MOE Key Laboratory for Nonequilibrium Synthesis and Modulation of Condensed Matter, School of Physics, Xi'an Jiaotong University, Xi'an 710049, China}
\author{Xing Wang} 
\affiliation{MOE Key Laboratory for Nonequilibrium Synthesis and Modulation of Condensed Matter, School of Physics, Xi'an Jiaotong University, Xi'an 710049, China}
\author{Dieter H. H. Hoffmann}
\affiliation{MOE Key Laboratory for Nonequilibrium Synthesis and Modulation of Condensed Matter, School of Physics, Xi'an Jiaotong University, Xi'an 710049, China}
\author{Fei Gao}
\affiliation{School of Physics and Optoelectronic Technology, Dalian University of Technology, Dalian 116024, China}
\author{Zhanghu Hu}
\affiliation{School of Physics and Optoelectronic Technology, Dalian University of Technology, Dalian 116024, China}
\author{Younian Wang}
\affiliation{School of Physics and Optoelectronic Technology, Dalian University of Technology, Dalian 116024, China}
\author{Wei Yu}
\affiliation{State Key Laboratory of High Field Laser Physics, 
Shanghai Institute of Optics and Fine Mechanics, 201800 Shanghai, China}
\author{Stephan Fritzsche}
\affiliation{Helmholtz-Institut-Jena and 
Friedrich-Schiller-University-Jena, D-07743 Jena, Germany}
\author{Xiantu He}
\affiliation{Key Laboratory of HEDP of the Ministry of Education and CAPT, 
Peking University, 100871 Beijing, China}
\date{\today}
\begin{abstract}
We compare the transport properties of a well-characterized hydrogen plasma for low and high current ion beams. The energy-loss of low current beams can be well understood, within the framework of current stopping power models. However, for high current proton beams, significant energy-loss reduction and collimation is observed in the experiment. We have developed a new particle-in-cell code, which includes both collective electromagnetic effects and collisional interactions. Our simulations indicate that resistive magnetic fields, induced by the transport of an intense proton beam, act to collimate the proton beam and simultaneously deplete the local plasma density along the beam path. This in turn causes the energy-loss reduction detected in the experiment.   
  
\end{abstract}

\pacs{52.38.Kd, 41.75.Jv, 52.35.Mw, 52.59.-f}

\maketitle

\noindent
\textbf{Introduction}

New phenomena in stopping and transport properties of plasma for intense ion beams are permanently discovered and reveal new insight into the details of the underlying basic physics \cite{Ren2020}. This is not only of fundamental interest but it is also essential for a number of applications like nuclear-fusion \cite{Edwards2013} and advanced accelerator techniques \cite{Sharkov2016}. Ion stopping in matter also has a number of applications in medicine \cite{Linz2007,Linz2016}, material science \cite{Pelka2010,Feldman2017,Zylstra2015}, and the determination of equation of state properties \cite{Pelka2010,Feldman2017}. Interaction of ion beams with solids and gases has been studied to a great extent during last century, since swift ions became available as fission products and later by the fast development of accelerator technology. The high standard of present-day detector technology in nuclear physics, particle physics and medicine is not conceivable without the enormous amount of research that has been invested into research on ion stopping in matter. Compared to the vast amount of data that exists for the stopping power of gaseous and solid matter under ambient conditions, the amount of data available for ionized matter and matter under extreme conditions is still scarce.  However, the situation is improving since a number of groups worldwide started to address the problem with plasma targets and laser generated particle beams, as well as ion beams from accelerators. But there are still open questions. This is due to the transient nature of plasma which poses a challenge to experiments. Moreover, the theoretical characterization is also cumbersome. When the beam current is high, collective electromagnetic effects in addition to the effects of binary close and distant collisions need to be taken into account. After the initial experiments in the 1980s \cite{Young1982,Olsen1985,Evans1988}, systematic beam-plasma interaction experiments followed with different ion species ranging from protons to uranium and plasma targets generated by low-density discharges, as well high-density Z-pinch plasma and laser generated plasma \cite{Weyrich1989,Deutsch1989,Hoffmann1990,Dietrich1992,Chabot1995,Belyaev1996,Wetzler1997,Golubev2001,Sakumi2001,Frank2013,Zylstra2015,Chen2020}. The ion beam density in all these experiments was low, and therefore beam-induced collective electromagnetic fields were negligible. Instead, it was of paramount interest to study the effects of plasma parameters like temperature, free electron density and ionization degree, on the ion stopping process. Due to the rapid development of laser technology, intense and high energy ion beams from laser acceleration processes are available \cite{Ren2020,Zylstra2015,Macchi2013}. High current beams (tens of kilo-amperes) can be focused to tens of micrometers \cite{Bartal2012}. Such ion beams, when interacting with plasmas, can significantly influence the plasma state, through self-generated electromagnetic fields and close interactions. Recently, the electromagnetic effects for proton beams’ transport in plasmas have been reported by using hybrid particle-in-cell (PIC) simulations \cite{Kim2015,Kim2016}. Strong focusing of proton beams was found in their simulations, and the effect was attributed to the generation of resistive magnetic fields. In this work, we cover our experiments on stopping and transport properties of ion beams with varying currents in hydrogen plasmas. Surprisingly, for high current proton beams, significant collimation and energy loss reduction is observed. In order to analyze the experimental results, a new particle-in- cell code (PIC code) was developed, which includes both collective electromagnetic effects and collisional interactions. The simulation suggests that self-generated electromagnetic fields induced by high current beams lead to focusing and in some cases to reduced energy loss.

\vspace{2ex}
\noindent
\textbf{Experimental setup and results}

The experiment was performed at the 320 kV high-voltage platform of HIRFL (Heavy Ion Research Facility at Lanzhou) \cite{Zhao2009,Cheng2018,Cheng2018_2,Zhao2020}, as shown in Fig. \ref{fig1}(a). An ion source of the electron cyclotron resonance (ECR) type (see section I of the Supplementary information for details) can generate positively charged ions in a wide range of species ranging from protons and alpha particles up to Ca. The energy range is limited by the maximum terminal voltage of 320 kV and the charge state of the ion charge state q. In our experiment, helium ion beams with an initial energy of 100 keV/u and proton beams in the energy range of 100 keV to 250 keV are produced. For the proton beam, the current is on the order of mA, which is several orders of magnitude higher than the beam current used in previous beam-plasma transport experiments \cite{Neff2006,Boggasch1992}. The helium ion beams have a very low current and are shot for comparison.
A gas discharge, especially designed for this experiment served as the plasma target. (see section II of the information for details). Detailed investigation of this plasma, following the procedures described in Ref. \cite{Kuznetsov2013} showed that it is very reproducible, stable and homogenous. After passing through the plasma and a bending magnet, the ion beam was then recorded by a fast-gated position-sensitive-detector (PSD). 

\begin{figure}[htb]
	\includegraphics[width=8.50cm]{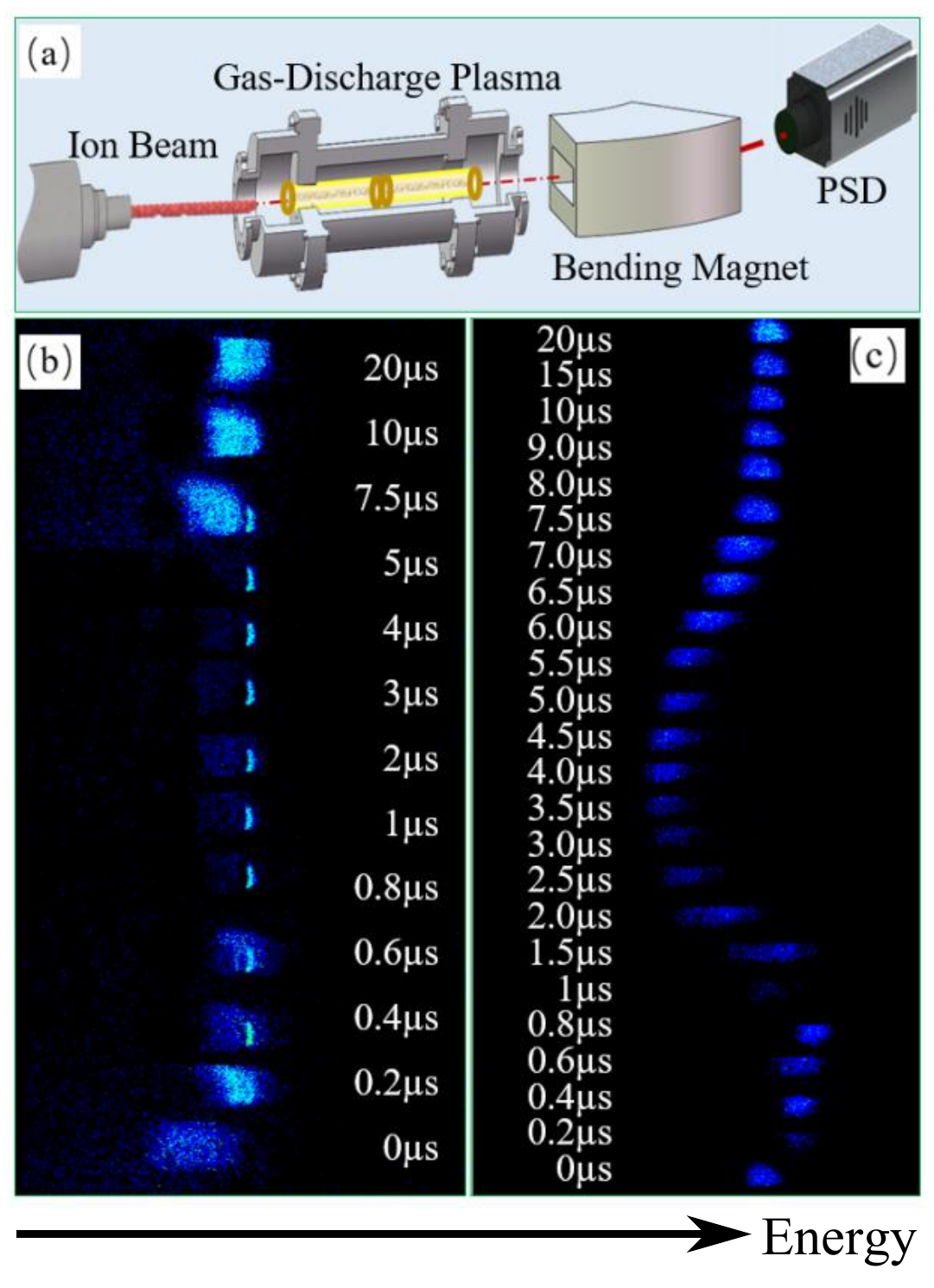}
	\caption{\label{fig1}(color online) 
		\textbf{Layout of the experiment and PSD results. }
		\textbf{a}  Schematic representation of the experimental facility, including ion source, hydrogen plasma target, bending magnet and fast-gated position-sensitive-detector.
		\textbf{b} Experimental energy losses (represented by position shifts) of high current proton beam in plasmas as a function of time, where the injected energy of protons is 200 keV.
		\textbf{c} Experimental energy losses of low current helium beam in plasmas as a function of time, where the injected energy of helium ions is 400 keV.}
\end{figure}

In Fig. \ref{fig1}(b), we show the results of an energy loss measurement of a high current (mA) proton beam with initial energy of 200 keV, after passing through the plasma. A position sensitive detector records the emitted ions at different times with respect to the plasma discharge, and the energy loss is represented by the shift in position. The energy loss depends on the time-varying free electron density of the hydrogen discharge plasma. Surprisingly, we observe a strong collimation effect for protons. Moreover, the energy loss of the collimated protons is significantly reduced. This collimation effect and the reduction in energy loss appears within a time window from 0.4 to 7.5 $\mu$s, when most of the hydrogen plasma has the maximum ionization degree. Within this time window, the maximum energy of the proton beam is 197.3 keV. This amounts to an energy loss of only 2.7 keV during the transport. This is significantly lower than the energy loss in the cold hydrogen gas before the plasma discharge. All previous energy loss measurements in discharge plasma and z-pinch plasma reported an enhancement of the energy loss. Thus, we are confronted with a new phenomenon.
Fig. \ref{fig1}(c) shows the experimental energy loss for helium ions. Our measurements agree very well with theoretical predictions, as reported in Ref. \cite{Zhao2021}.

The analysis of our experimental result displayed in Fig. \ref{fig1}(b) and (c), leads us to the following conclusion. 1) the collimation effect tends to occur at high plasma density; 2) the energy loss is significantly reduced for collimated ion beams. This is a new effect and as concluded above. Therefore, a new ansatz is necessary and we will explain the result by a simulation that takes into account collective effects.

\vspace{2ex}
\noindent
\textbf{Discussion}

In order to uncover the reason for the observed result, we here refer to particle in cell (PIC) simulations, since here collective effects like electromagnetic fields and plasma movement can well be included. To connect with the existing standard stopping power model, in PIC simulations, close particle collisions are included by a recently developed Monte-Carlo binary collision model \cite{Wu2017,Wu2017_2}. However, for typical PIC simulations \cite{Langdon1973}, the plasma frequency needs to be resolved, and moreover the grid size must be comparable to the Debye length in order to minimize artificial grid heating and to suppress numerical instabilities. Therefore, high noise levels and high computational requirements due to the operation on the shortest time and length scales greatly limit the applications of PIC methods to large scale simulations.

\begin{figure}[htb]
	\includegraphics[width=8.50cm]{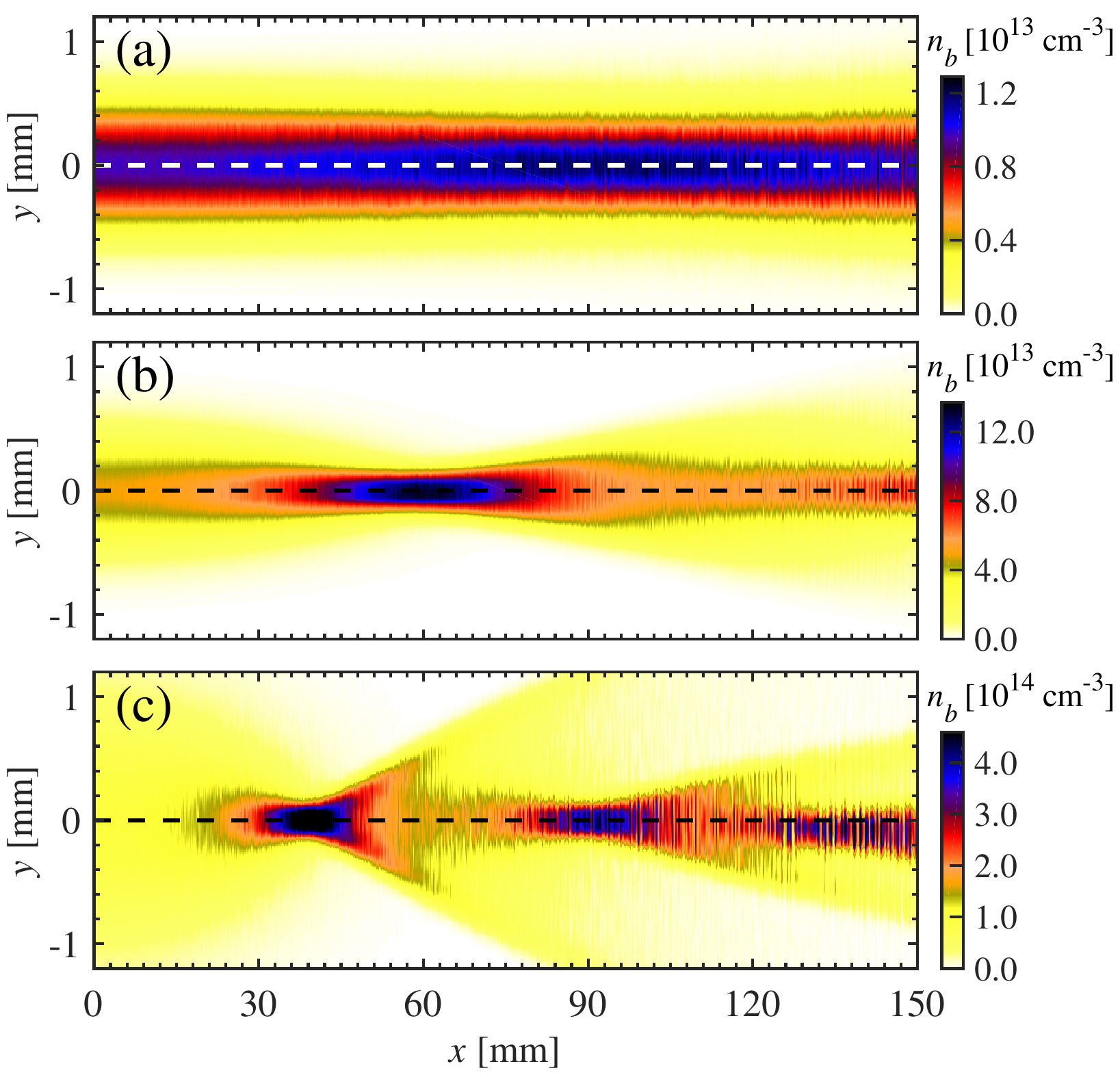}
	\caption{\label{fig2}(color online) 
		\textbf{The transport of proton beams in plasmas at 50 ns, obtained by PIC simulations, with different initial densities but the same injected energy 200 keV.} Sub-figures \textbf{a}, \textbf{b} and \textbf{c} correspond to initial densities of 1$\times$10$^{13}$, 
		5$\times$10$^{13}$ and 1$\times$10$^{14}$ cm$^{-3}$, respectively.}
\end{figure}

\begin{figure}[htb]
	\includegraphics[width=8.5cm]{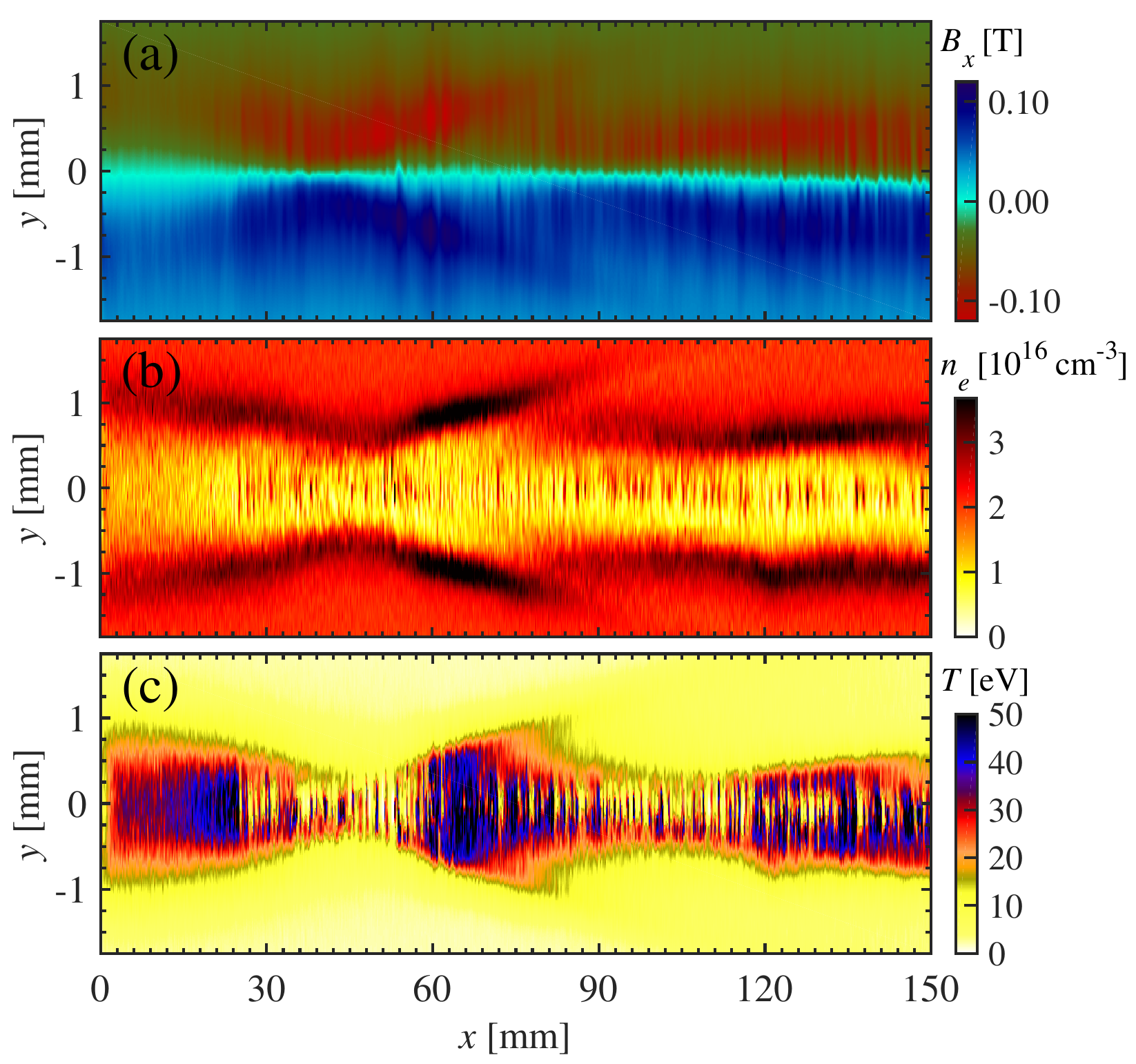}
	\caption{\label{fig3}(color online) 
		\textbf{PIC simulations for proton beam with density 1$\times$10$^{14}$ cm$^{-3}$.} The spacial distributions of \textbf{a} magnetic field, \textbf{b} plasma density, and \textbf{c} plasma temperature at 50 ns are presented. The injected energy of beam is 200 keV.
	}
\end{figure}

\begin{figure}[htb]
	\includegraphics[width=8.5cm]{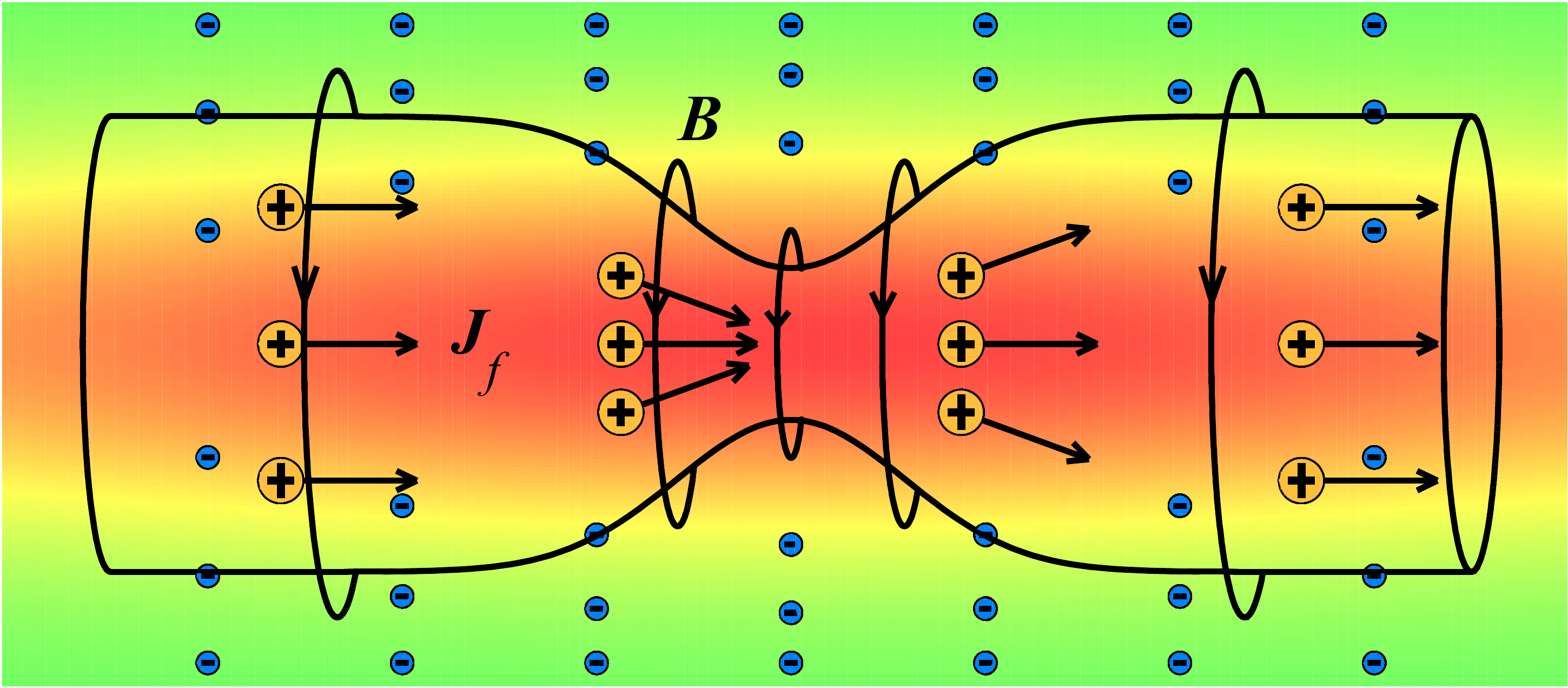}
	\caption{\label{fig4}(color online) 
		\textbf{Schematic representation of the plasma channel.} The yellow positively charged particles are the beam protons, and the blue negatively charged particles stand for plasma electrons. The initial plasma temperature is 1 eV (green). The beam can heat up the electrons in the center (red), and form a plasma channel.
	}
\end{figure}

Even though the beam current in the mA range is high, the beam particle density is still several orders of magnitude lower than the plasma density. Nevertheless we take into account electromagnetic effects induced by the beam. For this particular case, instead of solving the full Maxwell’s equations, we couple Ampere’s law, Faraday’s law and Ohm’s law, to update electric and magnetic fields. For a detailed description of this method, please refer to reference \cite{Wu2019}. As only a part of Maxwell’s equations needs to be solved, this method is of high speed, which is useful for large scale simulations. By using the two-dimensional LAPINS PIC code \cite{Wu2018,Wu2019_2}, the simulations were carried out in  X-Y Cartesian geometry with the ion beam propagating in the X direction. The size of the simulation box is chosen to be X(180 mm)$\times$Y(10 mm), which is divided into 1800$\times$100 uniform grids. The simulation assumes a uniform hydrogen plasma of 150 mm in length, free electron density $n_{fe}$=2.1$\times$10$^{16}$ cm$^{-3}$, bound electron density $n_{be}$=1.8$\times10^{16}$ cm$^{-3}$ and initial temperature $T_e$=1 eV. Proton beams with fixed energy 200 keV, FWHM (full width at half maximum) 1.4 keV and varying beam densities, 1$\times$10$^{13}$, 5$\times$10$^{13}$ and 1$\times$10$^{14}$ cm$^{-3}$ (corresponding to current densities, 1$\times$10$^{3}$, 5$\times$10$^{3}$ and 1$\times$10$^{4}$ A/cm$^{2}$), are injected into the plasma target. Fig. \ref{fig2} shows the beam distributions for different initial beam densities at 50 ns when propagating in the plasma target. The initial beam densities are (a) 1$\times$10$^{13}$, (b) 5$\times$10$^{13}$ and (c) 1$\times$10$^{14}$ cm$^{-3}$, respectively. With increasing beam density the injected beam starts collimating. The collimation mechanism can be well understood from Fig. \ref{fig3}. Here, the spacial distribution of (a) magnetic field (b) plasma density and (c) plasma temperature for beam with density 1$\times$10$^{14}$ cm$^{-3}$ at 50 ns are displayed. The co-moving background electrons (the so-called ``return current'') give rise to an electric and magnetic field as required by Maxwell's equations (Ohm's law and Faraday's law). Fig. \ref{fig3}(a) shows that the strength of magnetic field can be as high as 0.1 T, which is strong enough to collimate the injected proton beam. In Fig. \ref{fig3}(b), as time increases, a plasma channel is formed, which results from the heating of central background electrons in Fig. \ref{fig3}(c). A more intuitive illustration of this process is revealed in Fig. \ref{fig4}.

\begin{figure}[htb]
	\includegraphics[width=8.50cm]{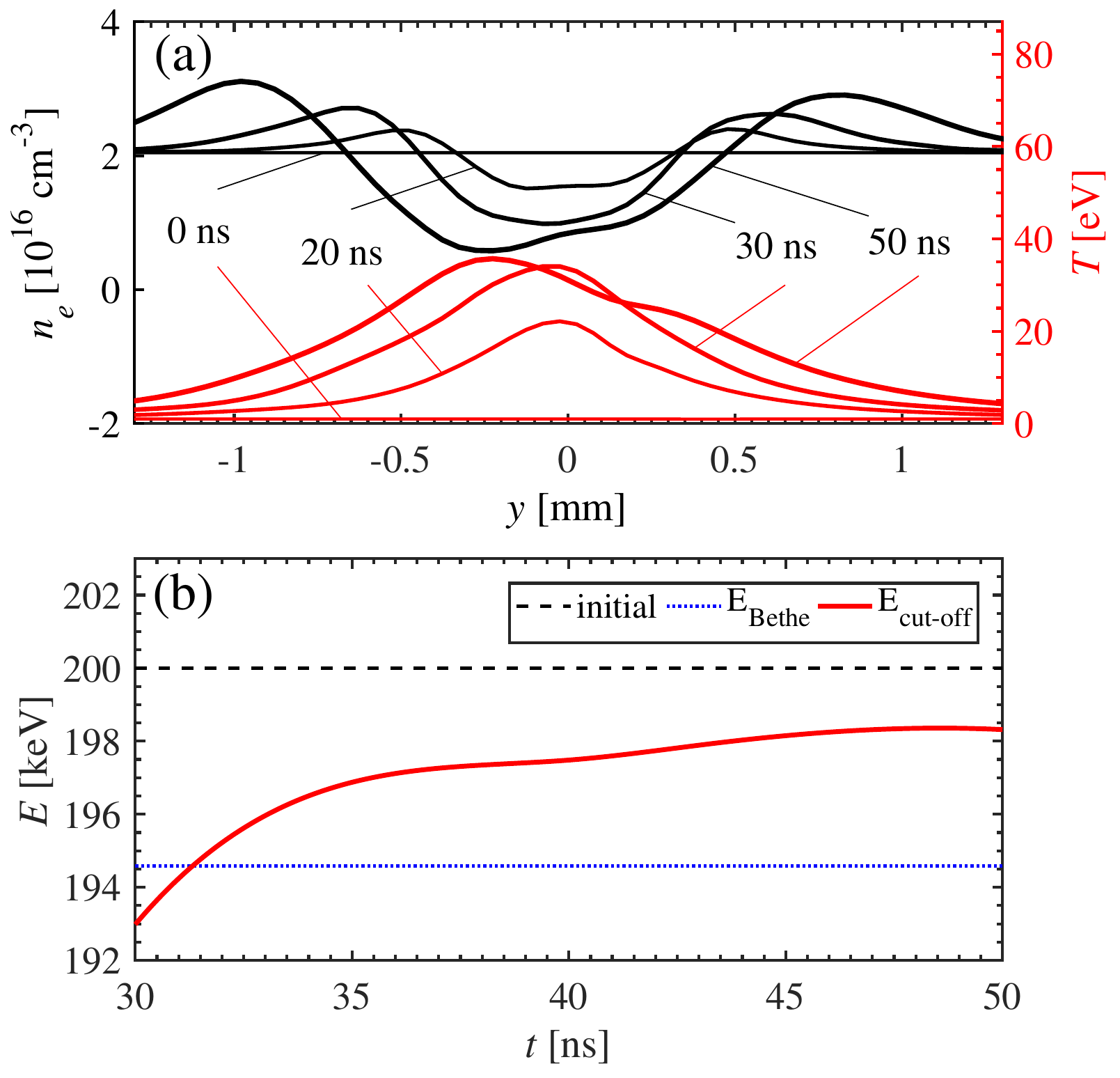}
	\caption{\label{fig5}(color online) 
		\textbf{The energy-loss reduction when beam gets collimated in plasmas.}
		\textbf{a} Profiles of electron density (black) and temperature (red) in $y$ direction at 0, 20, 30 and 50 ns.
		\textbf{b} Energy of ejected protons as a function of time. The solid curve is the cut-off energy in the simulation (red). The dotted line at $E$=194.6 keV is the result calculated by Bethe theory (blue). For comparison, the data for the incident 200-keV proton beam (black) are shown as well.
	}
\end{figure}

In Fig. \ref{fig5}(a), we have displayed how the profiles of electron density and temperature change with time.
From 0 to 50 ns, the central electrons are heated and the density of these particles drops.
With the increasing of simulation time, a plasma channel is forming. 
When collimated proton beams propagate through this low density plasma channel, 
their energy losses are reduced. 
The cut-off energy $E_{cut-off}$ of ejected beam protons, as a function of time, is displayed in Fig. \ref{fig5}(b). It stands for the higher energy value corresponding to half of the maximum intensity in energy spectrum. The blue dotted line at $E$ = 194.6 keV illustrates the result given by Bethe theory \cite{Bethe1930,Bethe1932} (see section IV of the Supplementary information for details). 
The cut-off energy in the simulation (solid red) grows with time and finialy exceed the predictions of Bethe theory. 

Up to now, we have reproduced the beam collimation and corresponding energy-loss reduction by using the PIC simulations. 
Note, the magnetic fields are proven to be the key factors. Combing Ampere's law, Ohm's law and Faraday's law, 
the magnetic fields can be expressed approximately as $\partial \bm{B}/\partial t=c\nabla\times{(\eta\bm{J}_f)}$. 
By increasing the beam current $\bm{J}_f$ and the resistivity $\eta$, the magnetic fields can be increased accordingly. 
This mechanism explains the collimation effect and the energy-loss reduction  for intense beams in the plasma channel. It also explains why such phenomena occur when the ionization of plasma target is high, since the resistivity $\eta$ depends linearly on the plasma ionization degree $Z$.  

To summarize, we report experimental results of transport properties of ion beams with varying currents in hydrogen plasmas. With fixed plasma conditions, the transport properties of low current helium ion beam and high current proton beam are compared with each other.  The energy-loss of low current beams can be well understood by current theoretical models. In contrast, for the high current proton beams, significant collimation and energy-loss reduction are found when passing through the plasmas. This result is corroborated by two-dimensional PIC simulations, which include both collective electromagnetic effects and collision interactions. We attribute collimation effects to the resistive magnetic fields induced by the transport of high current proton beams, and the energy-loss reduction is caused by the decrease of local plasma density along the beam path. 

\vspace{2ex}
\noindent
\textbf{Methods}

\noindent
\textbf{PIC simulation model.}
Simulations are performed by using the two-dimensional LAPINS PIC code \cite{Wu2018,Wu2019_2}, where close interactions including proton-nuclei, proton-bound electron, proton-free electron were treated
by a Monte Carlo binary collision model and an ionization dynamics model, including collisional ionization, electron-ion recombination and ionization potential depression, was also considered. Simulation of large scale plasmas often results in an intractable burden on computer power. Therefore, instead of solving
the full Maxwell’s equations, we used a new approach by combining the PIC method with a reduced model \cite{Wu2019_2}. To take into account collective electromagnetic effects, the background electron inertia is neglected. The generations of electric field and magnetic field are due to Ohm's law and Faraday's law, 
which are caused by the co-moving background electrons. 
The background ``return current'', $\bm{J}_b$, is calculated by the Ampere's law,
with $\bm{J}_b=(c/4\pi)\nabla\times\bm{B}-(1/4\pi)\partial\bm{E}/\partial t-\bm{J}_f$, where $\bm{J}_f$ is the current of injected protons.
The electric fields are then solved by the Ohm's law, $\bm{E}=\eta\bm{J}_b-\bm{v}_{e}\times\bm{B}/c-\nabla p_{e}/(en_{e})$, where $\eta$ represents the resistivity, $\bm{v}_{e}$ is the electron velocity and $p_{e}$ is the electron pressure.
Finally, Faraday's law is used to advance the magnetic fields $\partial \bm{B}/\partial t=-c\nabla\times\bm{E}$. Compared with the standard PIC code, this code is more time-saving and can
avoid numerical instabilities \cite{Wu2019_2}.

\vspace{2ex}
\noindent
\textbf{Data availability}
\\
\noindent
The datasets generated and analyzed during the current study are available from the corresponding authors upon reasonable request. The simulation details are available from the corresponding author on reasonable request.


\begin{thebibliography}{99} 

\bibitem{Ren2020} J. Ren, Z. Deng, W. Qi, et al., Observation of a high degree of stopping for laser-accelerated intense proton beams in dense ionized matter, Nat. Commun. 11, 5157 (2020).
\bibitem{Edwards2013} M. J. Edwards, P. K. Patel, J. D. Lindl, et al., Progress towards ignition on the national ignition facility, Phys. Plasmas 20, 070501 (2013).
\bibitem{Sharkov2016} B. Y. Sharkov, D. H. H. Hoffmann, A. A. Golubev, and Y. Zhao, High energy density physics with intense ion beams, Matter Radiat. Extremes 1, 28 (2016).
\bibitem{Linz2007} U. Linz and J. Alonso, What will it take for laser driven proton accelerators to be applied to tumor therapy?, Phys. Rev. Spec. Top.-Accel. Beams 10, 094801 (2007).
\bibitem{Linz2016} U. Linz and J. Alonso, Laser-driven ion accelerators for tumor therapy revisited, Phys. Rev. Accel. Beams 19, 124802 (2016).
\bibitem{Pelka2010} A. Pelka, G. Gregori, D. O. Gericke, et al., Ultrafast melting of carbon induced by intense proton beams, Phys. Rev. Lett. 105, 265701 (2010).
\bibitem{Feldman2017} S. Feldman, G. Dyer, D. Kuk, and T. Ditmire, Measurement of the equation of state of solid-density copper heated with laser-accelerated protons, Phys. Rev. E 95, 031201 (2017).
\bibitem{Zylstra2015} A. B. Zylstra, J. A. Frenje, P. E. Grabowski, et al., Measurement of charged-particle stopping in warm dense plasma, Phys. Rev. Lett. 114, 215002 (2015).
\bibitem{Young1982} F. C. Young, D. Mosher, S. J. Stephanakis, et al., Measurements of enhanced stopping of 1-mev deuterons in target-ablation plasmas, Phys. Rev. Lett. 49, 549 (1982).
\bibitem{Olsen1985} J. N. Olsen, T. A. Mehlhorn, J. Maenchen, and D. J. Johnson, Enhanced ion stopping powers in high-temperature targets, J. Appl. Phys. 58, 2958 (1985).
\bibitem{Evans1988} P. M. Evans, A. P. Fews, and W. T. Toner, Diagnosis of laser-produced plasmas using fusion reaction-products, Laser Part. Beams 6, 353 (1988).
\bibitem{Weyrich1989} K. Weyrich, D. H. H. Hoffmann, J. Jacoby, et al., Energy-loss of heavy-ions in a hydrogen discharge plasma, Nucl. Instrum. Methods Phys. Res., Sect. A 278, 52 (1989).
\bibitem{Deutsch1989} C. Deutsch, G. Maynard, R. Bimbot, et al., Ion beam-plasma interaction - a standard model approach, Nucl. Instrum. Methods Phys. Res., Sect. A 278, 38 (1989).
\bibitem{Hoffmann1990} D. H. H. Hoffmann, K. Weyrich, H. Wahl, et al., Energy-loss of heavy-ions in a plasma target, Phys. Rev. A 42, 2313 (1990).
\bibitem{Dietrich1992} K. G. Dietrich, D. H. H. Hoffmann, E. Boggasch, et al., Charge state of fast heavy-ions in a hydrogen plasma, Phys. Rev. Lett. 69, 3623 (1992).
\bibitem{Chabot1995} M. Chabot, D. Gardes, J. Kiener, et al., Charge-state distributions of chlorine ions interacting with cold gas and with fully ionized plasma, Laser Part. Beams 13, 293 (1995).
\bibitem{Belyaev1996} G. Belyaev, M. Basko, A. Cherkasov, et al., Measurement of the coulomb energy loss by fast protons in a plasma target, Phys. Rev. E 53, 2701 (1996).
\bibitem{Wetzler1997} H. Wetzler, W. Suss, C. Stockl, et al., Density diagnostics of an argon plasma by heavy ion beams and spectroscopy, Laser Part. Beams 15, 449 (1997).
\bibitem{Golubev2001} A. Golubev, V. Turtikov, A. Fertman, et al., Experimental investigation of the effective charge state of ions in beam-plasma interaction, Nucl. Instrum. Methods Phys. Res., Sect. A 464, 247 (2001).
\bibitem{Sakumi2001} A. Sakumi, K. Shibata, R. Sato, et al., Energy dependence of the stopping power of mev o-16 ions in a laser-produced plasma, Nucl. Instrum. Methods Phys. Res., Sect. A 464, 231 (2001).
\bibitem{Frank2013} A. Frank, A. Blazevic, V. Bagnoud, et al., Energy loss and charge transfer of argon in a laser-generated carbon plasma, Phys. Rev. Lett. 110, 115001 (2013).
\bibitem{Chen2020} B. Z. Chen, D. Wu, J. R. Ren, D. H. H. Hoffmann, and Y. T. Zhao, Transport of intense particle beams in large-scale plasmas, Phys. Rev. E 101, 051203 (2020).
\bibitem{Macchi2013} A. Macchi, M. Borghesi, and M. Passoni, Ion acceleration by superintense laser-plasma interaction, Rev. Mod. Phys. 85, 751 (2013).
\bibitem{Bartal2012} T. Bartal, M. E. Foord, C. Bellei, et al., Focusing of short-pulse high-intensity laser-accelerated proton beams, Nat. Phys. 8, 139 (2012).
\bibitem{Kim2015} J. Kim, B. Qiao, C. McGuffey, M. S. Wei, P. E. Grabowski, and F. N. Beg, Self-consistent simulation of transport and energy deposition of intense laser-accelerated proton beams in solid-density matter, Phys. Rev. Lett. 115, 054801 (2015).
\bibitem{Kim2016} J. Kim, C. McGuffey, B. Qiao, et al., Varying stopping and self-focusing of intense proton beams as they heat solid density matter, Phys. Plasmas 23, 043104 (2016).
\bibitem{Zhao2009} Y. Zhao, G. Xiao, H. Xu, et al., An outlook of heavy ion driven plasma research at imp-lanzhou, Nucl. Instrum. Methods Phys. Res., Sect. B 267, 163 (2009).
\bibitem{Cheng2018} R. Cheng, X. Zhou, and Y. Wang, Energy loss of protons in hydrogen plasma, Laser Part. Beams 36, 98 (2018).
\bibitem{Cheng2018_2} R. Cheng, Y. Lei, and X. Zhou, Warm dense matter research at hiaf, Matter Radiat. Extremes 3, 85 (2018).
\bibitem{Zhao2020} Y. Zhao, Z. Zhang, R. Cheng, et al., High-energy-density physics based on hiaf, Sci. Sin.-Phys. Mech. Astron. 50, 112004 (2020).
\bibitem{Neff2006} S. Neff, R. Knobloch, D. Hoffmann, et al., Transport of heavy-ion beams in a 1 m free-standing plasma channel, Laser Part. Beams 24, 71 (2006).
\bibitem{Boggasch1992} E. Boggasch, A. Tauschwitz, H. Wahl, et al., Plasma lens fine focusing of heavy-ion beams, Appl. Phys. Lett. 60, 2475 (1992).
\bibitem{Kuznetsov2013} A. P. Kuznetsov, O. A. Byalkovskii, R. O. Gavrilin, et al., Measurements of the electron density and degree of plasma ionization in a plasma target based on a linear electric discharge in hydrogen, Plasma Phys. Rep. 39,
248 (2013).
\bibitem{Zhao2021} Y. T. Zhao, Y. N. Zhang, R. Cheng, B. He, C. L. Liu, X. M. Zhou, Y. Lei, Y. Y. Wang, J. R. Ren, X. Wang, Y. H. Chen, G. Q. Xiao, S. M. Savin, R. Gavrilin, A. A. Golubev, and D. H. H. Hoffmann, Benchmark experiment to prove the role of projectile excited states upon the ion stopping in plasmas, Phys. Rev. Lett. 126, 115001 (2021).
\bibitem{Wu2017} D. Wu, X. T. He, W. Yu, and S. Fritzsche, Monte carlo approach to calculate ionization dynamics of hot soliddensity plasmas within particle-in-cell simulations, Phys. Rev. E 95, 023208 (2017).
\bibitem{Wu2017_2} D. Wu, X. T. He, W. Yu, and S. Fritzsche, Monte carlo approach to calculate proton stopping in warm dense matter within particle-in-cell simulations, Phys. Rev. E 95, 023207 (2017).
\bibitem{Langdon1973} A. B. Langdon, Energy-conserving plasma simulation algorithms, J. Comput. Phys. 12, 247 (1973).
\bibitem{Wu2019} D. Wu, W. Yu, Y. T. Zhao, D. H. H. Hoffmann, S. Fritzsche, and X. T. He, Particle-in-cell simulation of
transport and energy deposition of intense proton beams in solid-state materials, Phys. Rev. E 100, 013208 (2019).
\bibitem{Wu2018} D. Wu, X. T. He, W. Yu, and S. Fritzsche, Particle-in-cell simulations of laser-plasma interactions at solid densities and relativistic intensities: the role of atomic processes, High Power Laser Sci. Eng. 6, e50 (2018).
\bibitem{Wu2019_2} D. Wu, W. Yu, S. Fritzsche, and X. T. He, High-order implicit particle-in-cell method for plasma simulations at solid densities, Phys. Rev. E 100, 013207 (2019).
\bibitem{Bethe1930} H. Bethe, On the theory of the passage of rapid charged particle radiation through matter, Ann. Phys. (Leipzig) 5, 325 (1930).
\bibitem{Bethe1932} H. Bethe, About the side radiation and the nature of the coloring matter in natural blue halite, Z. Phys. 76, 293 (1932).

\end{thebibliography}

\vspace{2ex}
\noindent
\textbf{Acknowledgements}
\\
\noindent
This work was supported by National Key R$\&$D Program of China (Grant No. 2019YFA0404900), Science Challenge Project (Grant No. TZ2016005), National Natural Science Foundation of China (Grants No. 11705141, No. 11605269, No. 11775282, No. 11875096, No. 11775042, and No. U1532263), and China Postdoctoral Science Foundation (Grants No. 2017M623145 and No. 2018M643613).

\vspace{2ex}
\noindent
\textbf{Author contributions}
\\
\noindent
Yongtao Zhao conceived this work, organized the experiments and simulations with Rui Cheng and Dong Wu, respectively. The experiment was carried out by Yongtao Zhao, Rui Cheng, Xianming Zhou, Yu Lei, Yuyu Wang, and Jieru Ren, and Xing Wang. Benzheng Chen, Dong Wu, and Yongtao Zhao performed the simulations and related analysis. Guoqing Xiao, Xing Wang, Dieter Hoffmann, Fei Gao, Zhanghu Hu, Younian Wang, Wei Yu, Xin Qi, Stephan Fritzsche, and Xiantu He contribute in the physical discussion. Dong Wu, Yongtao Zhao, and Benzheng Chen wrote the paper.

\vspace{2ex}
\noindent
\textbf{Competing interests}
\\
\noindent
The authors declare no competing interests.

\end{document}